\documentclass[aps,pra,twocolumn,longbibliography]{revtex4-1}
\usepackage{graphicx}
\pdfoutput=1
\usepackage{dcolumn}
\usepackage{amsmath,bm}
\usepackage{color}

\begin{document}


\title{Dissipation-induced topological transitions in continuous Weyl materials}


\author{Kunal Shastri}
\affiliation{School of Electrical and Computer Engineering, Cornell University, Ithaca, NY 14853, USA}

\author{Francesco Monticone}  
\email[]{francesco.monticone@cornell.edu}
\affiliation{School of Electrical and Computer Engineering, Cornell University, Ithaca, NY 14853, USA}

\date{\today}

\begin{abstract}
Many topologically non-trivial systems have been recently realized using electromagnetic, acoustic, and other classical wave-based platforms. As the simplest class of three-dimensional topological systems, Weyl semimetals have attracted significant attention in this context. However, the robustness of the topological Weyl state in the presence of dissipation, which is common to most classical realizations, has not been studied in detail. In this paper, we demonstrate that the symmetry properties of the Weyl material play a crucial role in the annihilation of topological charges in the presence of losses. 
We consider the specific example of a continuous plasma medium and compare two possible realizations of a Weyl-point dispersion based on breaking time-reversal symmetry (reciprocity) or breaking inversion symmetry. We theoretically show that the topological state is fundamentally more robust against losses in the nonreciprocal realization. Our findings elucidate the impact of dissipation on three-dimensional topological materials and metamaterials.
 
\end{abstract}

\maketitle



\section{Introduction}

Weyl semi-metals are a new class of materials with linear degeneracies in the three-dimensional momentum space, called Weyl points, which carry a topological charge and are sources or sinks of Berry curvature. These properties make Weyl materials an important example of three-dimensional topological systems. Weyl points have been predicted and/or observed in various periodic systems including condensed-matter electronic systems \cite{lv_experimental_2015,huang_weyl_2015,xu_discovery_2015}, ultra-cold gases \cite{lan_dirac-weyl_2011,ganeshan_constructing_2015,dubcek_weyl_2015,zhang_simulating_2015,shastri_realizing_2017}, photonic \cite{lu_experimental_2015,lin_photonic_2016} and acoustic crystals \cite{chen_acoustic_2018,li_weyl_2018,ge_experimental_2018}. Recently, Weyl points have also been found to exist in continuum media with plasmonic dispersion \cite{xiao_hyperbolic_2016,gao_photonic_2016}. Various unusual physical properties can be found in these systems, including robust surface-wave states called Fermi Arcs \cite{wan_topological_2011}, anomalous magneto-resistance due to chiral anomaly  \cite{yu_predicted_2016,tchoumakov_magnetic-field-induced_2016,udagawa_field-selective_2016} and quantized circular photo-galvanic effects \cite{de_juan_quantized_2017}. Each of these unique properties holds promise for far-reaching applications in electronic, photonic, and acoustic systems.

It is well established that the Weyl semi-metal state is extremely robust to perturbations. This stems from the fact that Hermitian perturbations to the system's Hamiltonian can only move Weyl points around in momentum space, without changing the topology of the system. This can be seen from the effective Hamiltonian around a Weyl point, which is given by  $\mathcal{H}=\Sigma_i\alpha_ik_i\bm{\sigma_i}$, where $\bm{\sigma_i}$ are the Pauli matrices, $\alpha_i$ are scalar constants and $\bm{k}$ is the wavevector. Since all three Pauli matrices are used in the effective Weyl Hamiltonian $\mathcal{H}$ (all degrees of freedom are exhausted) \cite{armitage_weyl_2018}, adding a perturbation of the form $\mathcal{H}'=\Sigma_i\alpha_i'\bm{\sigma_i}$ will only change the location of the Weyl points but not destroy them. For a topological transition to occur, the system should be suitably modified such that two Weyl points of opposite charge overlap in momentum space. In principle, this charge annihilation can always be achieved since Weyl points come in pairs carrying equal and opposite topological charge, such that the Brillouin zone does not enclose any net charge. When they collide, the two Weyl points would dissipate their topological charge and transition to a topologically trivial state such as a gapped insulator or a Dirac semi-metal \cite{armitage_weyl_2018}. 

Recently, Cerjan et. al. \cite{cerjan_effects_2018, cerjan_experimental_2019} have predicted another means by which Weyl points can dissipate their topological charge. In the presence of a non-Hermitian perturbation such as absorption loss, a Weyl point transforms into a closed contour of exceptional points (called a Weyl exceptional ring) carrying the same topological charge. If any point on this Weyl exceptional ring comes into contact with a point on another Weyl exceptional ring carrying the opposite topological charge, the two rings would dissipate their charge and the system would undergo a topological transition to a trivial state even without opening a bandgap. Thus, any mechanism that induces a non-Hermitian perturbation could, in principle, destroy the topological nature of the Weyl semi-metal along with its useful properties. However, no study to date has quantitatively investigated the robustness of these topological properties to the effect of losses, despite the importance of this issue for practical applications. In this paper, we systematically address this fundamental question, focusing on non-Hermitian perturbations due to dissipative losses.   

Although losses can be ignored in electronic systems due to charge conservation, most photonic, plasmonic, and acoustic systems are, by nature, dissipative. This is particularly relevant for plasmonic-based realizations of topological systems, which are inherently dissipative due to electron-scattering losses and surface-collision damping (Landau damping \cite{Khurgin_2017,Khurgin_2019}). Yet, most previous studies on these classical Weyl materials have largely ignored dissipation, which may lead to inaccurate predictions, as discussed, in a different context, in Ref. \cite{Ali_2019_Bulk-Edge}. Specifically, dissipation cannot be ignored in systems where oppositely charged Weyl points are located near each other in momentum space, which is the case for systems that only weakly break time-reversal or inversion symmetry. The size of the resulting Weyl exceptional rings depend on the magnitude of the non-Hermitian perturbation, which is monotonically related to the dissipation in the system. One can then imagine different scenarios in which nearby Weyl exceptional rings may come into contact in the presence of losses if the rings lie on the same plane in momentum space, or avoid each other completely if they lie on distinct parallel planes, resulting in a topological transition in one case and no transition in the other case. This suggests that the impact of dissipation on the topological phase of a Weyl material can be qualitatively deduced from geometrical/symmetry arguments and need not depend on the particular realization. In the following, we start by presenting these general arguments in details, and then illustrate the most relevant results by considering a continuous plasmonic medium, with broken time-reversal or inversion symmetry, as a model system of dissipative Weyl materials for electromagnetic waves. 

\section{Results}

\subsection{The role of symmetries}
\label{sec:minimal_model}

\begin{figure*}
\includegraphics[scale=0.65]{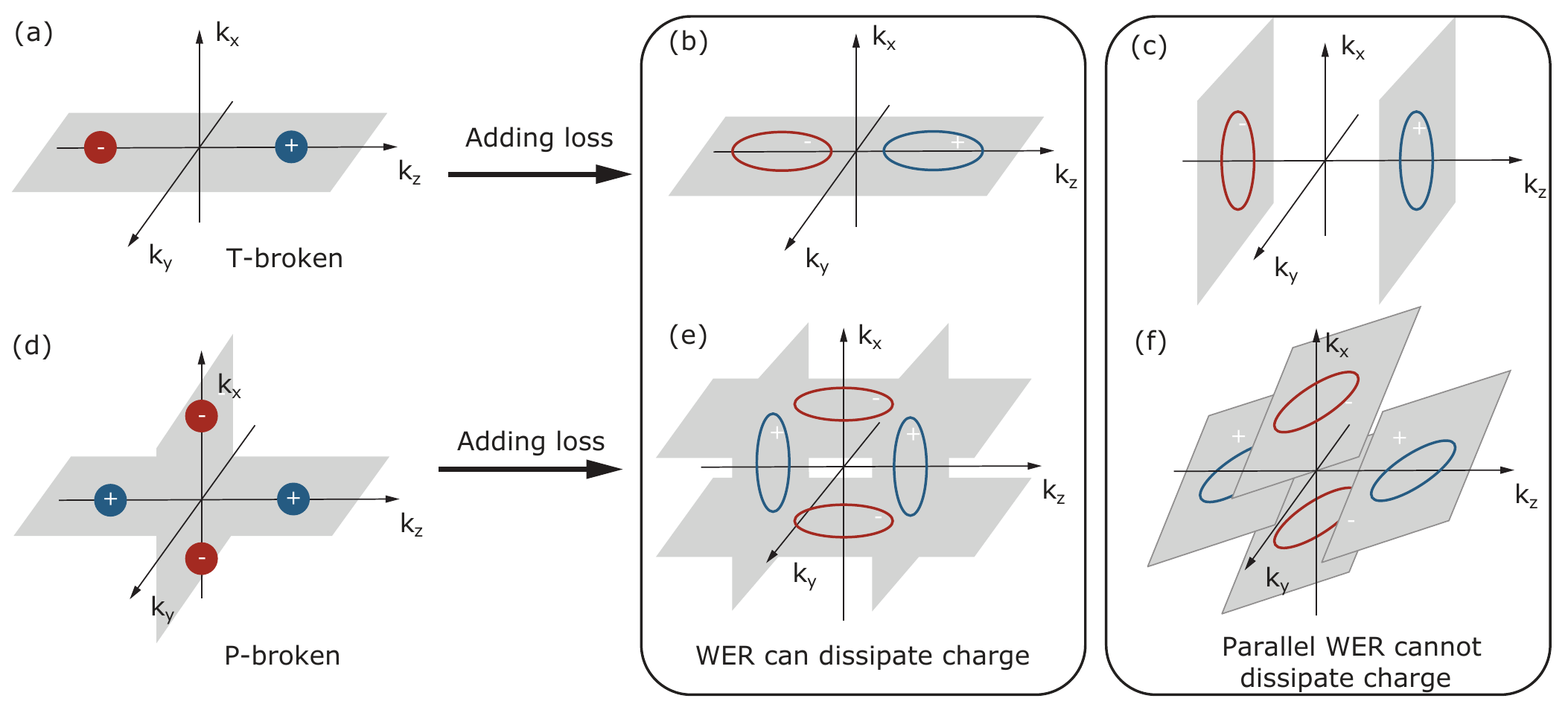}
\caption{Weyl points and exceptional rings. (a) Illustration of two charge-1 Weyl points in a time-reversal-broken (nonreciprocal) Weyl material. (b,c) Any small non-Hermitian perturbation, such as absorption loss, will transform the points into Weyl exceptional rings (WER) with the same topological charge, which may lie on the same plane (b) or on distinct parallel planes (c). (d) Illustration of the minimum number of charge-1 Weyl points in an inversion-symmetry-broken (chiral) Weyl material, and (e,f) two possible configurations of the corresponding exceptional rings in the presence of loss. Dissipation of the topological charge is only possible if oppositely charged exceptional rings come into contact as loss is increased.}\label{fig:0}
\end{figure*}

In this section we consider two separate cases, a Weyl material that breaks time-reversal symmetry but preserves inversion symmetry and one that preserves time reversal and breaks inversion. The first case will have a minimum of two inversion-related Weyl points with equal and opposite topological charge, as shown in Fig \ref{fig:0}(a) (applying the inversion operator to one of the two Weyl points yields the other Weyl point with opposite charge). In general, for a non-Hermitian perturbation that is isotropic and independent of the wavevector (local), a Weyl point carrying unit topological charge is transformed into a planar ring of exceptional points \cite{cerjan_effects_2018}. Thus, a topological transition would only be possible if the two Weyl exceptional rings of opposite charge lie on the same plane or on intersecting planes, as illustrated in Fig \ref{fig:0}(b,c). For the time-reversal-broken case, we show in the following that the rings are located on distinct parallel planes and the presence of losses does not typically result in a topological transition. We also would like to note that time-reversal symmetry is equivalent to Lorentz reciprocity in the lossless case; hence, when referring to the non-perturbed Hamiltonian of our systems we will use these concepts interchangeably. Dissipation alone would break time-reversal symmetry, but not reciprocity. 

Consider a minimal two-level Hamiltonian $\mathcal{H}_0(\bm{k})=k_x\bm{\sigma_x}+k_y\bm{\sigma_y}+(k_z^2-\alpha)\bm{\sigma_z}$ describing two Weyl points located at $\bm{k_\pm}=(0,0,\pm\sqrt{\alpha})$ with charge $\pm1$ as shown in Fig. \ref{fig:0}(a). Suppose then that the inversion operator is given by $P=\sigma_z$ and the  time-reversal operator by $T=U\mathcal{K}$, where $\mathcal{K}$ is the anti-linear complex conjugation operator and $U$ is a unitary operator. It can be seen that the above Hamiltonian preserves inversion symmetry $P^\dagger \mathcal{H}_0(\bm{-k})P=\mathcal{H}_0(\bm{k})$ but breaks time-reversal symmetry $T^\dagger \mathcal{H}_0(\bm{-k})T\neq \mathcal{H}_0(\bm{k})$. To model dissipative loss in the system, we then include a generic non-Hermitian perturbation term given by $\mathcal{H}_1=j\Sigma_ig_i\bm{\sigma_i}$ that does not depend on the wavevector $\bm{k}$. Let us also assume that this perturbation preserves the inversion symmetry of the system $P^\dagger \mathcal{H}_1P=\mathcal{H}_1$. In this case, $\mathcal{H}_1$ will reduce to the form $\mathcal{H}_1=jg_z\bm{\sigma_z}$ since $\bm{\sigma_{x,y}}$ are odd under inversion. It follows that the dispersion of the eigenvalues of the perturbed Hamiltonian $\mathcal{H}_0(\bm{k})+jg_z\bm{\sigma_z}$ consists of two exceptional circular rings with radius $\sqrt{g_z}$, having the same topological charge as the Weyl point, located on \emph{distinct} parallel planes along the $k_z$ axis, as shown in Fig. \ref{fig:0}(c). Thus, irrespective of the magnitude of $g_z$, or the location of the unperturbed Weyl points on the $k_z$ axis, the pair of Weyl exceptional rings of opposite topological charge will not come into contact. This argument will also apply to a Weyl material containing more than one pair of Weyl exceptional rings, provided that the annihilating pairs are related by inversion symmetry. 

The assumption that the non-Hermitian perturbation does not break any additional symmetries that are preserved by the unperturbed Hamiltonian, except time reversal, is crucial to our argument above. This is typically true in the case of classical wave-physics systems where the non-Hermitian contribution is from material dissipative losses, which may be inhomogeneous and anisotropic, but are unlikely, per se, to break inversion (parity) symmetry. Moreover, we would like to note that, in realistic systems, large losses may result in Hermitian perturbations as well, which would move the Weyl exceptional rings in momentum space and may induce topological transitions, as further discussed in the next sections. Furthermore if the non-Hermitian perturbation is nonlocal (spatially dispersive), namely, it depends on the wavevector $\bm{k}$, then the coefficients $g_{x,y}(\bm{k})$ in the non-Hermitian perturbation term need not be zero, potentially resulting in Weyl exceptional rings located on the same plane, as illustrated in Fig. \ref{fig:0}(b). 

Let us now consider the second scenario of interest, that is, a reciprocal Weyl material that breaks inversion symmetry [Fig. \ref{fig:0}(d,e,f)]. In this case, time-reversal symmetry dictates that pairs of Weyl points located at $\bm{k}$ and $\bm{-k}$ carry the same topological charge (applying the time-reversal operator to one of the two Weyl points yields the other Weyl point with the same charge). Hence, in the presence of loss, the resulting Weyl exceptional rings would not be able to annihilate each other since they have the same charge. However, different from the previous case, here the minimum number of Weyl points is four, since other two Weyl points having opposite charge need to be present so that the total net charge is zero \cite{armitage_weyl_2018} (this is strictly true only for periodic systems; for continuum systems, for which the momentum space is not compact, this statement is still true if nonlocal effects are properly included in the form of a high spatial-frequency cutoff for the material response, as done in Ref. \cite{silveirinha_chern_2015}). Thus, in this case, each ring would have to touch a second pair of Weyl exceptional rings with opposite charge to cause a topological transition to a trivial state. However, since these two pairs of oppositely charged Weyl exceptional rings are not generally related by symmetry, we cannot conclude that they will remain parallel and not intersect. In this case, additional symmetry arguments specific to the Weyl material realization will be required to conclude whether the resulting Weyl exceptional rings, in the presence of loss, would be located on parallel or intersecting planes, as illustrated in Fig. \ref{fig:0}(e,f). As an example, in Fig. \ref{fig:0}(f), Weyl exceptional rings located on oblique parallel planes would violate mirror symmetry along the $x-y$ plane. If the dissipative system preserves this symmetry, then this Weyl ring arrangement is prohibited, and the rings will necessarily have to be located on intersecting planes. Hence, if the lossless Weyl points are located close to each other, dissipation may easily lead to a topological transition to a trivial state. 
 
These considerations hint at an intrinsic fragility of the topological phase of inversion-symmetry-broken Weyl materials in the presence of dissipation. Conversely, nonreciprocal Weyl materials appear fundamentally more robust.

\subsection{Weyl Points in Plasmonic Media}

\begin{figure}
\includegraphics[scale=0.65]{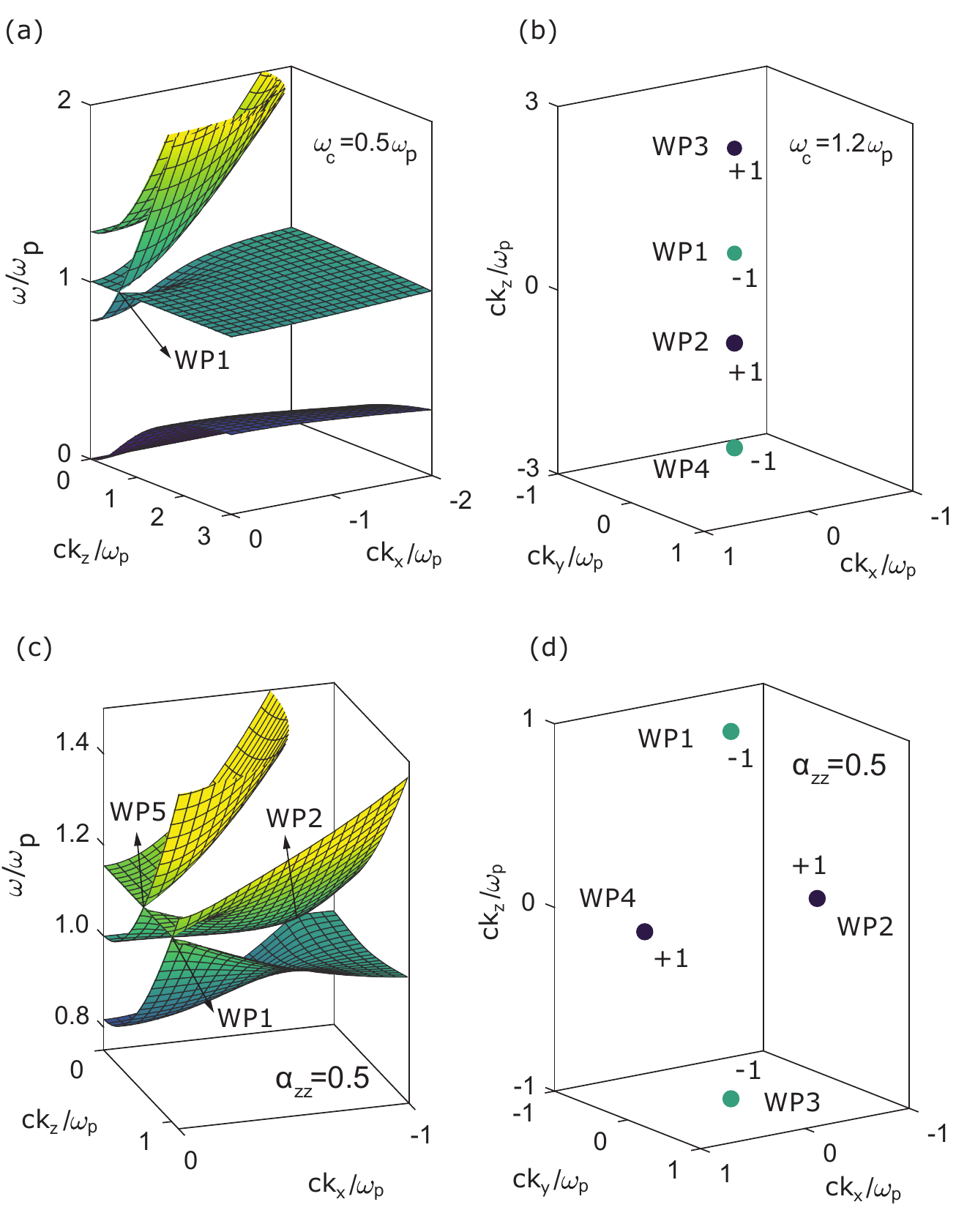}
\caption{Band diagrams and Weyl points in the non-dissipative scenario. (a) Modal dispersion surfaces for a lossless plasma magnetized along the $+z$ direction. The cyclotron frequency is $\omega_c=0.5\omega_p$, and $k_y=0$. The surface color corresponds to the frequency. (b) Illustration of all the Weyl points of a magnetized plasma, with $\omega_c=1.2\omega_p$, in three-dimensional $\bm{k}$ space. (c) Modal dispersion surfaces for a chiral plasmonic material with $\alpha_{zz}=0.5$, and $k_y=0$. (d) Location of the four Weyl points between the first and the second band in three-dimensional $\bm{k}$ space. The pair WP1-WP3 and the pair WP2-WP4 are at different frequencies.}\label{fig:1}
\end{figure}

In this section, we explore and clarify the general concepts discussed in the previous sections by considering a relevant physical implementation of an electromagnetic Weyl semi-metal: a continuous lossy plasma, or plasmonic (meta)material, with Weyl points \cite{gao_photonic_2016,xiao_hyperbolic_2016}. Specifically, we consider two possible ways of realizing Weyl-point dispersion in this model system: first, by breaking time-reversal symmetry (reciprocity) via an external magnetic bias, while preserving inversion symmetry; second, by breaking inversion symmetry via chiral coupling between the $z$ components of the electric and magnetic fields, while preserving reciprocity.  

The equation of motion for free charge carriers in a lossless plasmonic medium with plasma frequency $\omega_p$ can be written as $\omega_p^2\epsilon_{0}\bm{E}=j\omega\bm{J}$ for time-harmonic fields with $e^{j\omega t}$ temporal dependence. Here, $\bm{E}$ and $\bm{J}$ are the electric field vector and current density vector, and $\epsilon_0$ is the permittivity of free space. The dielectric response from bound charges is assumed to be zero, i.e., $\epsilon_\infty = 1$. We will comment on the effect of a larger than unity $\epsilon_\infty$ in the next section. Together with Maxwell's equations for plane waves, $\bm{k}\times\bm{E}/\mu_0 =\omega\bm{H}$ and $\bm{k}\times\bm{H}/\epsilon_0 =-\omega\bm{E}+j\bm{J}/\epsilon_0$, this can be rewritten in the form of an eigenvalue equation \cite{gao_photonic_2016, raman_photonic_2010}, $\bm{\mathcal{H}}\left |\psi_n\right\rangle=\omega_n/\omega_p\left |\psi_n\right\rangle$, where the eigenstates are $9\times1$ vectors of the form $\left |\psi_n\right\rangle=[\bm{E},\sqrt{\mu_0/\epsilon_{0}}\bm{H},\bm{J}/\omega_p\epsilon_{0}]^T$ and the Hamiltonian is given by,
\begin{equation} \label{eq:H0}
\bm{\mathcal{H}}(\bm{k})=\left( \begin{matrix}
\bm{0} & -c\bm{K}/\omega_p & j\bm{I} \\
c\bm{K}/\omega_p & \bm{0}& \bm{0} \\
-j\bm{I}  & \bm{0} & \bm{0}\\
\end{matrix} \right),
\end{equation}
where $\bm{K}$ is a $3\times3$ tensor given by $K_{ij}=k_k\varepsilon_{ikj}=(0,-k_z,k_y;k_z,0,-k_x;-k_y,k_x,0)$, where $\varepsilon_{ijk}$ is the Levi-Civita tensor ($\bm{K}$ is sometimes known as Kong's tensor in the electromagnetics literature). This system preserves both inversion and time-reversal symmetries. 

\subsubsection{Nonreciprocal implementation}

In the presence of an external static magnetic field, corresponding to cyclotron frequency $\omega_c$, time-reversal symmetry and reciprocity are broken. If the bias is applied along the $+\bm{z}$ direction, the Hamiltonian becomes, 

\begin{equation} \label{eq:HbT}
\bm{\mathcal{H}}(\bm{k})=\left( \begin{matrix}
0 & -c\bm{K}/\omega_p & j\bm{I} \\
c\bm{K}/\omega_p & \bm{0}& \bm{0} \\
-j\bm{I}  & \bm{0} & \omega_c\bm{\Delta}\\
\end{matrix} \right).
\end{equation}

Here, $\Delta_{ij}=z_k\varepsilon_{ikj}=(0,-1,0;1,0,0;0,0,0)$. Note that this Hamiltonian is a $9\times9$ Hermitian matrix. The eigen-frequency spectra (modal dispersion surfaces) are plotted in Fig \ref{fig:1}(a) for $k_y=0$ and $\omega_c=0.5\omega_p$. 

This magnetically biased plasmonic system has two regimes of interest.  When $0<\omega_c<\omega_p$, there are two Weyl points at $k_{\text{WP1,2}}=\pm\sqrt{\omega_c/(\omega_c+\omega_p)}$, with charge $-1$ and $+1$, respectively, corresponding to the linear degeneracies between the circularly polarized transverse mode and the longitudinal mode [Fig. \ref{fig:1}(a)]. This regime has been studied in detail in Refs. \cite{Ali_Unidirectional_2019,pakniyat_non-reciprocal_2020,yang_one-way_2016} If the bias is turned off, $\omega_c=0$, the system becomes reciprocal, the two Weyl points annihilate each other at $k_z=0$, and the Weyl material undergoes a topological phase transition to a trivial state. 

In the second regime of interest, when $\omega_c>\omega_p$, two more Weyl degeneracies appear at $k_{\text{WP3,4}}=\pm\sqrt{\omega_c/(\omega_c-\omega_p)}$, with charge $+1$ and $-1$, between the lower circularly polarized transverse mode and the longitudinal mode. As $\omega_c$ approaches $\omega_p$, these two Weyl points are created/annihilated at the Brillouin zone edge, i.e., $k=\pm \infty$ for a continuous system. Here, since the plasmonic medium is assumed continuous, the Brillouin zone extends to infinity in momentum space (the situation would be different for plasmonic metamaterials based on a periodic arrangement of meta-atoms). The properties of this second pair of Weyl points have been studied in \cite{gao_photonic_2016}.  The location of all the Weyl points for $\omega_c=1.2\omega_p$ in three-dimensional momentum space is shown in Fig \ref{fig:1}(b). 

Here, we shall focus on the Weyl-point pair occurring in the first regime, $0<\omega_c<\omega_p$, since this is a more common situation and easier to realize in practice due to the lower bias intensity (we also speculate that these type of Weyl points may be present in naturally occurring plasmas under weak magnetic bias, such as in atmospheric or astronomical scenarios).

\subsubsection{Chiral implementation}

For the realization of a reciprocal Weyl semi-metal based on a plasmonic material, following \cite{xiao_hyperbolic_2016}, we break inversion symmetry by including chiral coupling $(\alpha_{zz})$ between the $z$ components of electric and magnetic fields, and simultaneously including a non-unitary dielectric constant along the $y$ direction $(\epsilon_{ry})$. This system may be constructed in the form of a chiral wire-medium, i.e., a metamaterial consisting of helical (elliptic) wires in the $z$ direction \cite{xiao_hyperbolic_2016}, and straight wires in the other two directions. This system is mathematically modeled as an extended eigenvalue equation, $\bm{\mathcal{H}}\left |\psi_n\right\rangle=\omega_n/\omega_p\bm{C}\left |\psi_n\right\rangle$, where the tensor $\bm{\mathcal{H}}(\bm{k})$ is the same as in Eq. \eqref{eq:H0}, and the $9\times9$ tensor $\bm{C}$ is given by,

\begin{equation} \label{eq:HbI}
\bm{C}=\left( \begin{matrix}
\bm{\epsilon_r} & j\bm{\alpha}/c & \bm{0} \\
-j\bm{\alpha}/c & \bm{I}& \bm{0} \\
\bm{0}  & \bm{0} & \bm{I}\\
\end{matrix} \right),
\end{equation}
where $\bm{\epsilon_r}$ and $\bm{\alpha}$ are $diag(1,\epsilon_{ry},1)$ and $diag(0,0,\alpha_{zz})$, respectively, and $c$ is the speed of light in vacuum. The eigen-frequency spectra (modal dispersion surfaces) for this case are plotted in Fig \ref{fig:1}(c) for $k_y=0$, $\alpha_{zz}=0.5$ and $\epsilon_{ry}=1.5$.

As shown in Fig \ref{fig:1}(d), for $\alpha_{zz}>0$, this system has four Weyl points (WP1-4) between the first and second lowest frequency band for $\omega>0$. Additional Weyl points (WP5,6) are located between the second and third bands; however, these are not the focus of this work and will not be considered further. The four Weyl points (WP1-4) are located on the $k_y=0$ plane, with WP1 and WP3 on the $k_z$ axis at $k_z=\pm\omega_p\sqrt{(\alpha_{zz}^2+\epsilon_{ry}-1)/(1-\alpha_{zz}^2)}$ and frequency $\omega=\omega_p/\sqrt{1-\alpha_{zz}^2}$. Moreover, the pair WP1 and WP3 are related by time-reversal symmetry and, therefore, have the same topological charge of $+1$. The other time-reversal-related pair of Weyl points, WP2 and WP4, have a topological charge of $-1$ and are located on the $k_x$ axis at $k_x=\pm\omega_p\sqrt{(\epsilon_{ry}-1)(1+\sqrt{\delta})/2}$, where $\delta=(\epsilon_{ry}-1+4\alpha_{zz}^2)/(\epsilon_{ry}-1)$ and frequency  $\omega=\omega_p$. As expected, the two pairs of Weyl points are not at the same frequency since they are not related to each other by any symmetries. As the chiral coupling between the electric and magnetic fields, $\alpha_{zz}$, approaches zero, we verified that these four Weyl points, and the corresponding bands, merge into a single nodal ring, namely, a degeneracy curve in momentum space carrying zero topological charge.

\subsection{Weyl Exceptional rings}

\begin{figure*}
\includegraphics[scale=0.65]{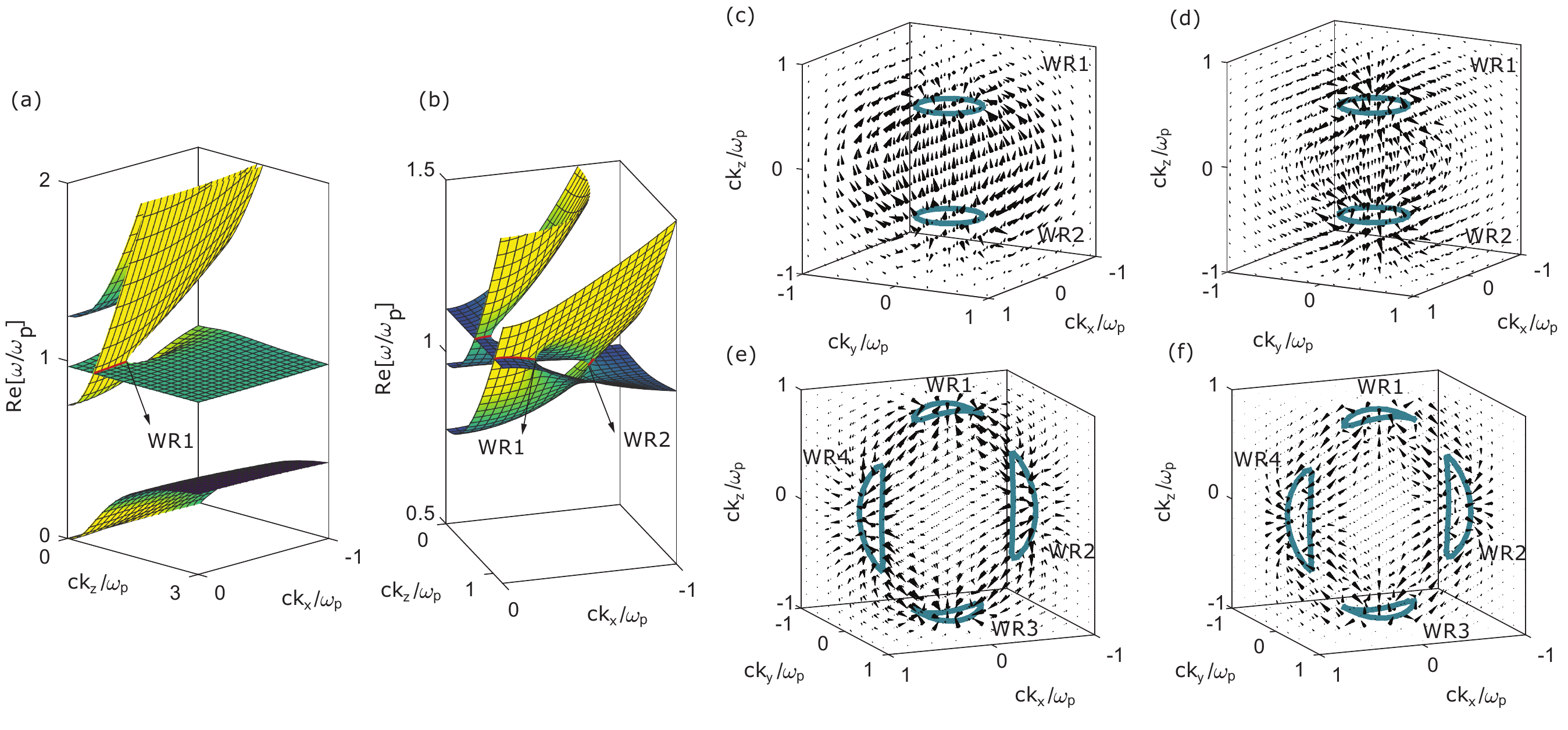}
\caption{Band diagrams and Weyl exceptional rings in the dissipative scenario. (a) Modal dispersion surfaces (real eigenfrequency spectrum) for a lossy magnetized plasma with $\omega_c=\gamma=0.5\omega_p$, and $k_y=0$. The imaginary component of the eigenfrequencies is indicated by the surface color of the bands. One of the Weyl exceptional rings, WR1, is clearly seen in the plot as an exceptional point in this two-dimensional momentum space. (b) Modal dispersion surfaces, similar to (a), but for a chiral plasmonic material with $\alpha_{zz}=0.5$, $\gamma=0.5\omega_p$, and $k_y=0$. In both (a) and (b), the bulk Fermi arcs (real-frequency degeneracies connecting the exceptional points) are highlighted in red. Real component (c) and imaginary component (d) of the Berry curvature for the lossy magnetized plasma. (e) and (f) are the corresponding plots for the lossy chiral plasmonic medium. In both cases, the imaginary component has zero divergence and does not contribute to topological charge calculations.}\label{fig:2}
\end{figure*}

Considering both the plasmonic Weyl systems discussed in the previous section, we now investigate how their dispersion properties and topology change in the presence of losses.

\subsubsection{Nonreciprocal implementation}

For the case of a magnetized plasma with scattering losses characterized by relaxation time $\tau$ and collision frequency $\gamma=1/\tau$, the Hamiltonian in Eq. \eqref{eq:HbT} becomes,

\begin{equation} \label{eq:HbTd}
\bm{\mathcal{H}}=\left( \begin{matrix}
0 & -c\bm{K}/\omega_p & j\gamma\bm{I}\\
c\bm{K}/\omega_p & 0& 0 \\
-j\gamma\bm{I}  & 0 & \omega_c\bm{\Delta}+j\gamma\bm{I}\\
\end{matrix} \right),
\end{equation}
where $\bm{I}$ is the identity matrix. The eigenfrequencies $\omega_n$ of this non-Hermitian Hamiltonian are complex and correspond to decaying modes. The real frequency spectra for dissipation $\gamma=0.5\omega_p$ and $k_y=0$ are plotted in Fig \ref{fig:2}(a), with the surface color scaling linearly with the imaginary component of the eigenfrequencies. In the two dimensional momentum space defined by $k_x$ and $k_z$, the two Weyl points, WP1 and WP2, located on the $k_z$ axis, split into two exceptional points each, with an offset from the $k_z$ axis. Each pair of exceptional points is connected by a real frequency line degeneracy, called bulk Fermi arc \cite{zhou_bulk-fermi_2018}, which is indicated by red lines in Fig. \ref{fig:2}(a,b). However, the imaginary parts of the eigenfrequencies are not degenerate along these lines, except at the exceptional points, as seen from the surface color in Fig \ref{fig:2}(a) (yellow for the transverse band and green for the longitudinal band). These exceptional points correspond to Weyl exceptional rings in three-dimensional momentum space. 

Following the discussion in Section \ref{sec:minimal_model}, if the dissipation term does not break parity symmetry, or the rotational symmetry around the direction of the magnetic bias $(\bm{z})$, the Weyl exceptional rings originating from Weyl points WP1 and WP2 lie on distinct parallel planes. We confirm this by plotting in Fig \ref{fig:2}(c) and (d) the exceptional rings at which the real and imaginary parts of the eigenfrequencies are degenerate in three-dimensional momentum space. 

In order to determine the total topological charge of these exceptional rings, we first calculate the complex Berry curvature of the frequency bands. Since the Hamiltonian in Eq. \eqref{eq:HbTd} is not Hermitian, the eigenstates $|\psi_n\rangle$ are not orthogonal \cite{brody_biorthogonal_2013}. However, as usually done for complex Hamiltonians, we can define a bi-orthogonal eigenbasis consisting of left $|\psi_n^l\rangle$ and right eigenstates $|\psi_n\rangle$ 
defined by the eigenvalue problems $\bm{\mathcal{H}}|\psi_n\rangle=\omega_n/\omega_p |\psi_n\rangle$ and $\bm{\mathcal{H}}^\dagger|\psi_n^l\rangle=\omega_n^*/\omega_p|\psi_n^l\rangle$. The Berry curvature can then be defined in four different ways based on a combination of left and right eigenvectors, but all four have been shown to result in the same topological invariant \cite{shen_topological_2018}. Here, we use the definition $\bm{\Omega}_n(\bm{k})= \nabla\times\langle\psi_n^l|\nabla|\psi_n\rangle$ to numerically calculate the complex Berry curvature of the non-Hermitian system. Since the derivative of the eigenstate in this formula introduces a gauge ambiguity in the numerical procedure, following the corresponding formula for Hermitian systems \cite{berry_quantal_1984}, we express the complex Berry curvature as a derivative of the Hamiltonian,

\begin{eqnarray} \label{eq:BC}
\Omega_{n,ij}(\bm{k})&=&\sum_{m\neq n}\frac{1}{(E_n-E_m)^2}\langle\psi_n^l|\frac{\partial \mathcal{H}}{\partial k_i}|\psi_m\rangle\langle\psi_m^l|\frac{\partial \mathcal{H}}{\partial k_j}|\psi_n\rangle\nonumber\\
&-&\langle\psi_n^l|\frac{\partial \mathcal{H}}{\partial k_j}|\psi_m\rangle\langle\psi_m^l|\frac{\partial \mathcal{H}}{\partial k_i}|\psi_n\rangle
\end{eqnarray}

Using this formula we then evaluate the real [Fig. \ref{fig:2}(c)] and imaginary [Fig. \ref{fig:2}(d)] components of the Berry curvature vector field for the lower band (with respect to real frequency) participating in the Weyl exceptional ring degeneracy. From these plots, it can be seen that the real-frequency-degeneracy surface enclosed by the Weyl exceptional rings WR2 and WR1 act, respectively, as a source or sink of real Berry curvature. This is expected since these two rings have emerged from Weyl points WP2 and WP1, respectively. In contrast, the imaginary component of the Berry curvature curls around the Weyl exceptional rings, analogous to the magnetic field lines around a current carrying loop. This implies that the imaginary component of the Berry curvature does not contribute to the closed surface integral in the topological charge calculations (its net flux is zero). 
Considering a cubic surface enclosing each ring individually, we numerically find that the topological charge of the Weyl rings is the same as for the corresponding Weyl points: $+1$ and $-1$ for WR2 and WR1, respectively. As the losses $\gamma$ in the system are reduced, the Weyl exceptional rings gradually shrink to a Weyl point and the imaginary Berry curvature vanishes, resulting in the usual monopole-like real Berry curvature field for Weyl points. 

\subsubsection{Chiral implementation}

We can similarly explore the effect of dissipation in a chiral plasmonic material by including the $j\gamma\bm{I}$ term in the Hamiltonian of Eq. \eqref{eq:H0} (as we have done in \eqref{eq:HbTd}) and substituting it into $\bm{\mathcal{H}}\left |\psi_n\right\rangle=\omega_n/\omega_p\bm{C}\left |\psi_n\right\rangle$, where $\bm{C}$ is given by Eq. \eqref{eq:HbI}. As an illustrative example, we set $\gamma=0.5\omega_p$ and plot the dispersion surfaces for $k_y=0$ in Fig \ref{fig:2}(b). Similar to the magnetically biased plasma case, all four Weyl points WP1-4 transform into Weyl exceptional rings WR1-4 in the presence of loss. Moreover, the disks enclosed by Weyl rings WR1 and WR3 are sources of real Berry curvature, whereas WR2 and WR4 are sinks, as shown in Fig \ref{fig:2}(e). In contrast, the imaginary component of the Berry curvature curls around rings, similar to the nonreciprocal case. 
By numerically integrating the Berry curvature we confirm that the topological charge is +1 for WR1 and WR3 and -1 for WR2 and WR4. A crucial difference from the magnetized plasma case is that the oppositely charged Weyl rings are no longer located on parallel planes; hence, they can easily annihilate each other if they come into contact as the dissipation in the system is increased or the chiral coupling term $\alpha_{zz}$ is reduced. Again, this fact hints at the intrinsic fragility of the topological phase of a chiral Weyl material, as further elucidated in the next section.

\subsection{Topological transitions}

\begin{figure}
\includegraphics[scale=0.68]{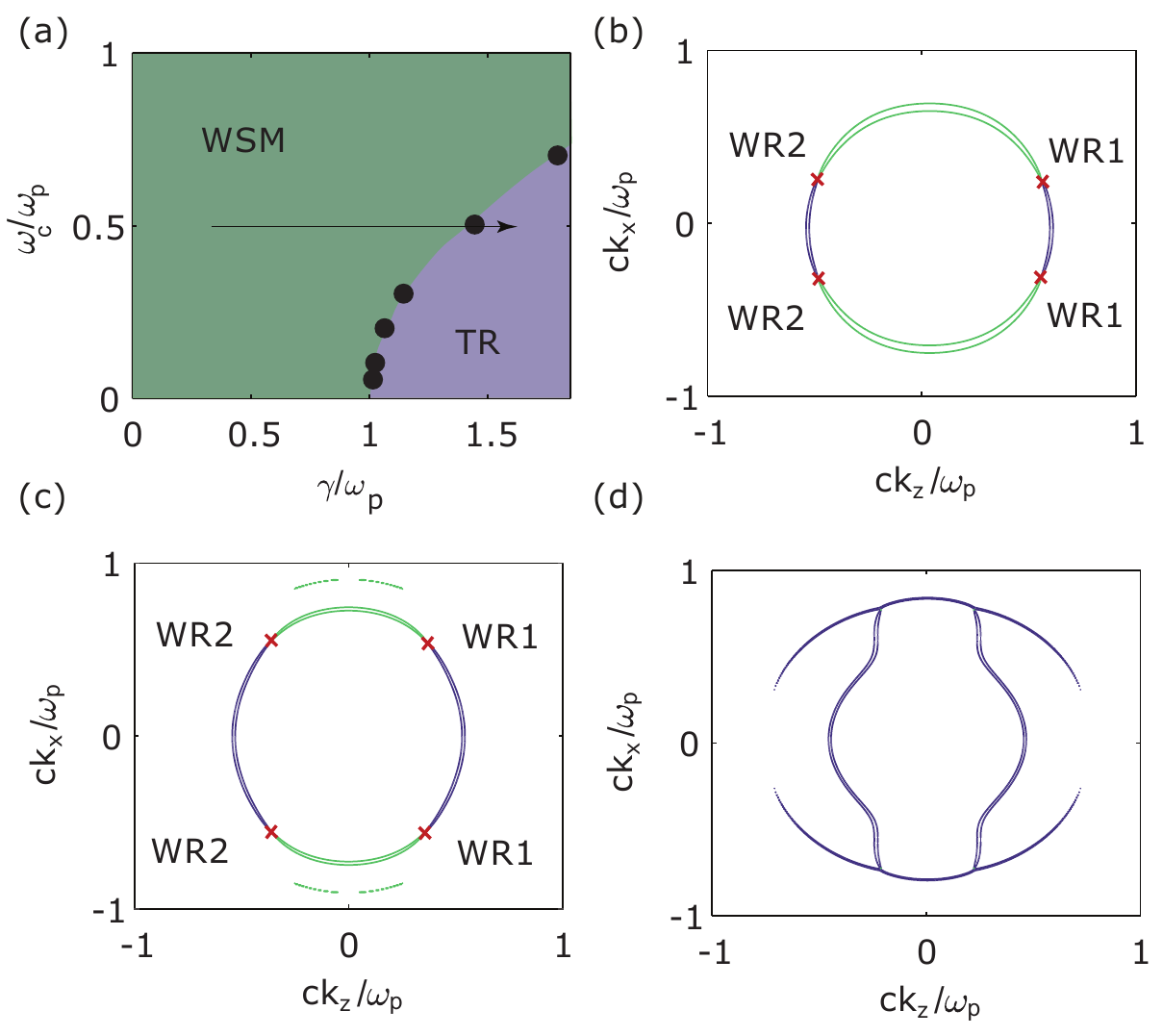}
\caption{(a) Phase diagram for a magnetized Weyl plasma with isotropic losses. Weyl semimetal phase and trivial phase are denoted by WSM and TR. (b)-(d) Plots of the modal degeneracies in two-dimensional momentum space for a plasma with $\omega_c=0.5\omega_p$. Purple and green curves indicate the degeneracy in the real and imaginary eigenfrequency spectrum, respectively. The intersections of real- and imaginary-frequency degeneracy surfaces correspond to Weyl exceptional rings (exceptional points in two dimensions, marked by solid red crosses). Loss rate is $\gamma=0.5\omega_p$ in (b), $\gamma=1.0\omega_p$ in (c) and $\gamma=1.5\omega_p$ in (d).}\label{fig:3}
\end{figure}

Since we are interested in studying the robustness of the topological Weyl state against dissipation, we track the location of the Weyl exceptional rings in the above cases for increasing levels of loss. Intuitively, if oppositely charged Weyl exceptional rings come into contact and merge, any closed surface surrounding this new contour of exceptional points would now enclose zero total topological charge (the net flux of Berry curvature is zero), corresponding to a trivial state. In other words, the topological charge has dissipated.

\subsubsection{Nonreciprocal implementation}

Starting with the nonreciprocal dissipative case, that is, a lossy magnetized plasma with Hamiltonian given by Eq. \eqref{eq:HbTd}, we numerically calculate the contours along which the relevant bands $(N+1,N)$ have a degeneracy $(\omega_{N+1}-\omega_N = 0)$ in momentum space. Consistent with our discussion in the previous section, for small non-zero losses, $\gamma$, the two Weyl point degeneracies WP1,2 satisfying $\operatorname{Re}[\omega_{N+1}-\omega_N]=0$ transform into two disk-like surfaces that are parallel to each other, corresponding to the purple curves in the two-dimensional momentum space in Fig. \ref{fig:3}(b)-(d). The degeneracy in the imaginary eigenfrequency spectrum satisfying $\operatorname{Im}[\omega_{N+1}-\omega_N]=0$ is a single surface that intersects the two real-frequency-degeneracy surfaces, corresponding to the green curves in Fig. \ref{fig:3}(b)-(d). The intersection of the two degeneracy surfaces correspond to the two Weyl exceptional rings WR1,2 indicated by solid blue curves in Fig. \ref{fig:2}(c),(d) and by red crosses in Fig \ref{fig:3}(b)-(d). Note that, in Fig. \ref{fig:3}, we have only plotted degeneracy surfaces in the $k_z$-$k_x$ plane since the dispersion is cylindrically symmetric around the $k_z$ axis. For a fixed cyclotron frequency $\omega_c=0.5\omega_p$, it can be seen in Figs. \ref{fig:3}(b) and (c) that, as the dissipation $\gamma$ is increased from $0.5\omega_p$ to $1.0\omega_p$, not only does the radius of the two Weyl exceptional rings (the distance between red crosses) increase, but they also get closer to each other along the $k_z$ axis. While the expansion of the Weyl rings is expected for a non-Hermitian perturbation, as discussed in Section \ref{sec:minimal_model}, their shift in momentum space is due to the fact that large losses also contribute to an Hermitian perturbations of the system's Hamiltonian, namely, the Hamiltonian in Eq. \eqref{eq:HbTd} cannot be simply written as an unperturbed Hermitian Hamiltonian plus a non-Hermitian perturbation. This behavior continues until finally, for $\gamma=1.5\omega_p$, the imaginary-frequency degeneracy surface disappears, as seen in Fig. \ref{fig:3}(d), and the two Weyl exceptional rings annihilate each other. We have numerically verified that the equal and opposite topological charge carried by the two exceptional rings cancel each other and the system undergoes a topological transition to a gap-less trivial state.

Increasing the cyclotron frequency increases the separation between the oppositely charged Weyl exceptional rings, thereby requiring higher levels of loss to undergo a topological transition. This follows from the lossless case, in which the separation between Weyl points WP1 an WP2 depends on the cyclotron frequency, as discussed above. To better illustrate this behavior, we map out the phase diagram for a dissipative magnetized plasma by numerically locating the critical loss $\gamma$ that results in a topological phase transition, for different values of cyclotron frequencies $\omega_c$. As expected, the phase map in Fig \ref{fig:3}(a) shows that increasing the cyclotron frequency monotonically increases the critical value of loss. Most importantly, for small cyclotron frequencies, and even in the limit $\omega_c \rightarrow 0$, $\gamma$ needs to be at least as large as the plasma frequency to change the topology of the system. Since such levels of loss are unrealistically high for most plasmas and solid-state plasmonic materials \cite{maier_plasmonics_2007}, we conclude that, for all practical purposes, dissipation cannot change the topology of a dissipative Weyl plasma. This resilience for arbitrarily low external magnetic fields may seem surprising, but can be explained by the relative location of the two exceptional rings on distinct parallel planes, which very slowly move toward each other as dissipation increases. As discussed in Section \ref{sec:minimal_model}, the exceptional rings would lie on intersecting planes only if the perturbation broke the inversion symmetry of the system.

\begin{figure}
\includegraphics[scale=0.65]{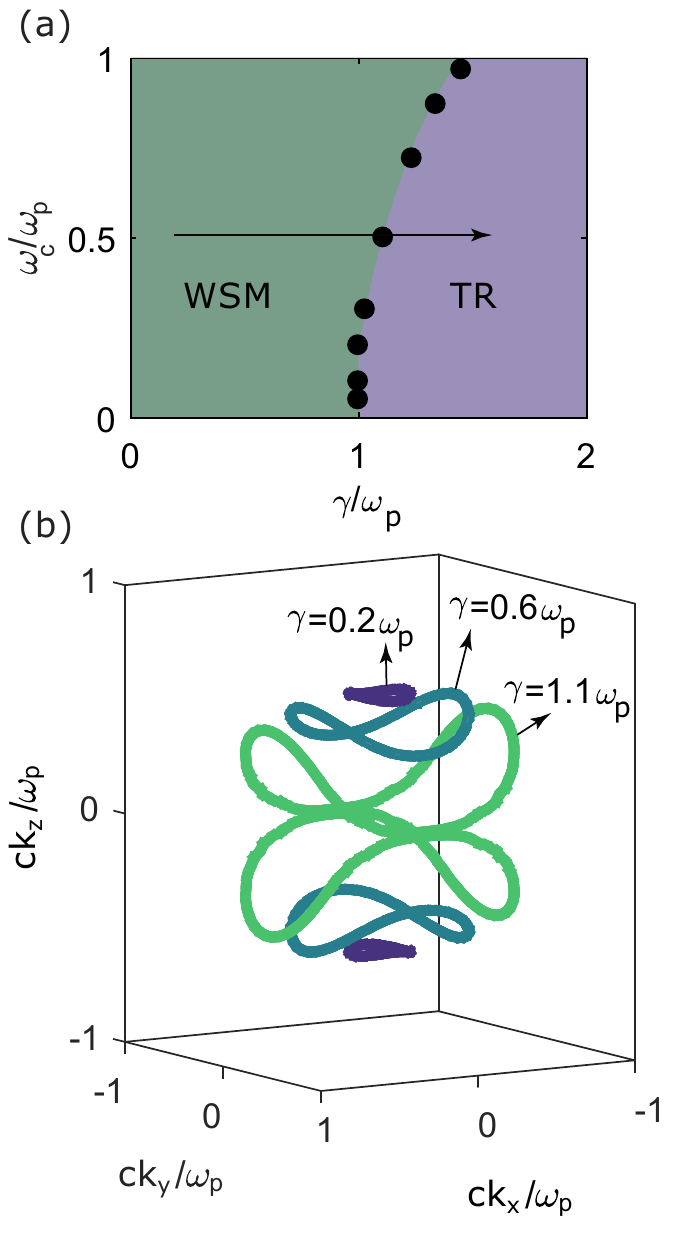}
\caption{(a) Phase diagram for a magnetized Weyl plasma with anisotropic losses (non-zero dissipation only along the $\bm{y}$ and $\bm{z}$ directions). (b) Weyl exceptional rings for different values of dissipation, $\gamma=0.2\omega_p, 0.6\omega_p$, and $1.1\omega_p$, and a fixed cyclotron frequency $\omega_c=0.5\omega_p$. It can be seen that the oppositely charged rings touch for $\gamma=1.1\omega_p$, corresponding to a topological phase transition as discussed in the text.}\label{fig:4}
\end{figure}

It may appear that the exceptional rings could be brought into contact at lower levels of dissipation if the rings could be made to \emph{bend} towards each other. Since this is different from having rings on intersecting planes, it does not require breaking inversion symmetry; instead, it simply requires a spatially anisotropic dissipation that breaks the rotational symmetry of the system about the $\bm{z}$ axis. As a relevant example, we consider a specific situation with large anisotropy, in which the dissipation occurs only along two directions, $\bm{y}$ and $\bm{z}$. This can be modeled by replacing the non-Hermitian dissipation term $j\gamma\bm{I}$ in Eq. \eqref{eq:HbT} with $j\gamma\bm{\Gamma}$ where the vector $\bm{\Gamma}$ is given by $[0,0,0;0,1,0;0,0,1]$. Either absorption losses or radiative losses in a strongly anisotropic metamaterial structure (e.g., layered metamaterials) could give rise to anisotropic losses, as the material resonances and loss mechanisms may be different for different polarization directions.  

For high cyclotron frequencies, it can be seen in Fig. \ref{fig:4}(a) that the numerically calculated phase diagram is different from the previous case with isotropic losses in Fig. \ref{fig:3}(a). In particular, at higher cyclotron frequencies, the topological transition can indeed occur at a lower critical value of loss. As shown in Fig. \ref{fig:4}(b), for $\omega_c=0.5\omega_p$, the Weyl exceptional rings for low dissipation, $\gamma=0.2\omega_p$,  look similar to the isotropic loss case in Fig. \ref{fig:2}(c,d). As the dissipation is increased, however, it can be seen that the two Weyl exceptional contours grow in size and curve towards each other along the $k_x=0$ plane. At $\gamma=1.1\omega_p$, the two contours lose their topological charge as they come into contact along the $k_z=0$ plane. This critical loss value is indeed smaller than in the isotropic loss case. However, for smaller cyclotron frequencies $\omega_c<0.5\omega_p$, the phase diagram looks very similar to the isotropic case. A minimum loss of $\gamma=\omega_p$ is still required for a topological transition. This is because the dissipation term only breaks the rotational symmetry but not the inversion symmetry of the system. Hence, it follows from the argument given in Section \ref{sec:minimal_model} that, for low levels of loss, the exceptional rings are located on distinct parallel planes independently of $\omega_c$, and a large level of perturbation is required for the exceptional rings to bend sufficiently to touch.

For the sake of completeness, we also note that the regime of cyclotron frequency higher than plasma frequency, $\omega_c>\omega_p$, results in two more Weyl degeneracies, WP3,4, between the first and the second band, as mentioned above and reported in \cite{gao_photonic_2016}. Compared to Weyl points WP1,2, these Weyl points and the corresponding Weyl rings for dissipative plasmas, are located further away in momentum space $k_{WP3,4}=\pm\sqrt{\omega_c/(\omega_c-\omega_p)}$. Thus, for these additional degeneracies, a topological transition induced by dissipation would occur for even larger losses. Finally, another plasma parameter that may affect the topological transition is the bound-charge dielectric constant $\epsilon_\infty$. In this work, we had set $\epsilon_\infty=1$ (no dielectric response from bound charges), and we have verified that increasing this constant monotonically moves the Weyl points WP1,2 further away from each other. 

All these findings confirm our observations that nonreciprocal Weyl states are inherently robust to the presence of dissipation.

\begin{figure}
\includegraphics[scale=0.65]{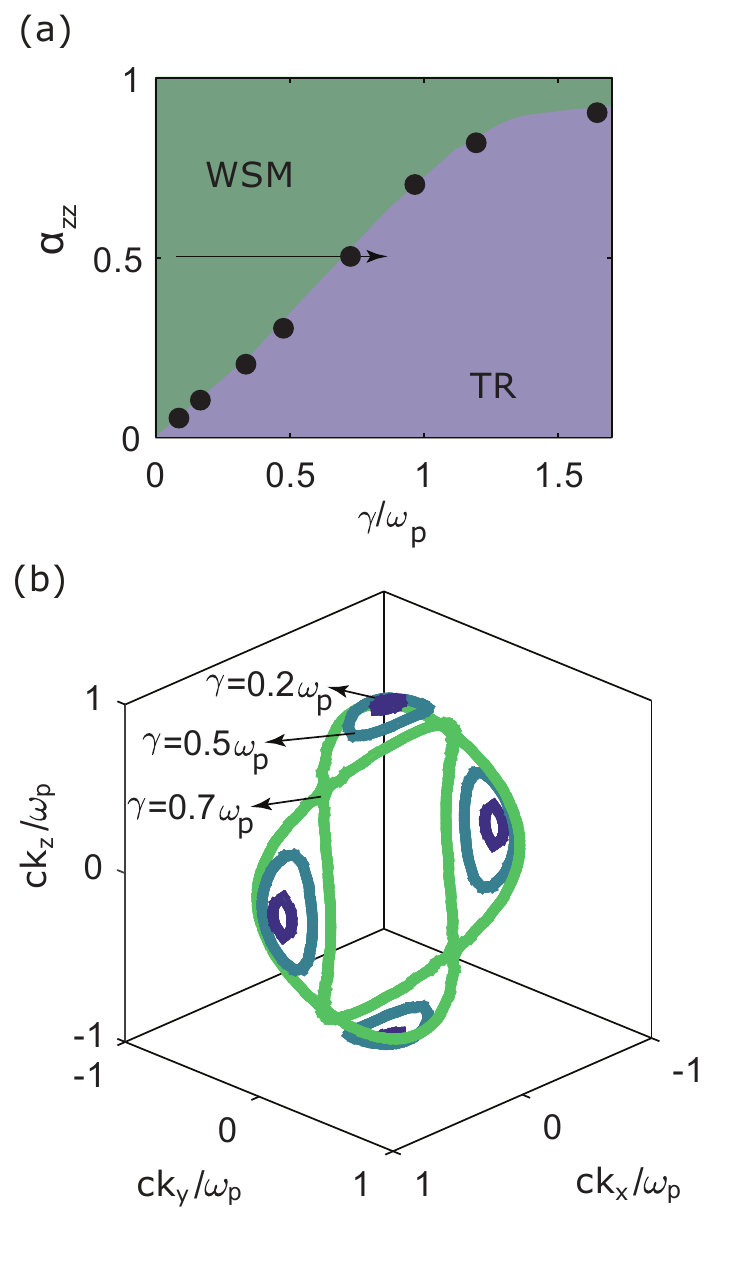}
\caption{(a) Phase diagram for a chiral Weyl plasma with isotropic losses. (b) Weyl exceptional rings for different values of dissipation, $\gamma=0.2\omega_p, 0.5\omega_p$, and $0.7\omega_p$, and a fixed chiral coupling between the $\bm{z}$ components of the electric and magnetic fields, $\alpha_{zz}=0.5$. An example of exceptional contour for losses larger than the critical value, and the corresponding Berry curvature, are shown in Fig. \ref{fig:6}.}\label{fig:5}
\end{figure} 

\subsubsection{Chiral implementation}

\begin{figure}
	\includegraphics[width=\columnwidth]{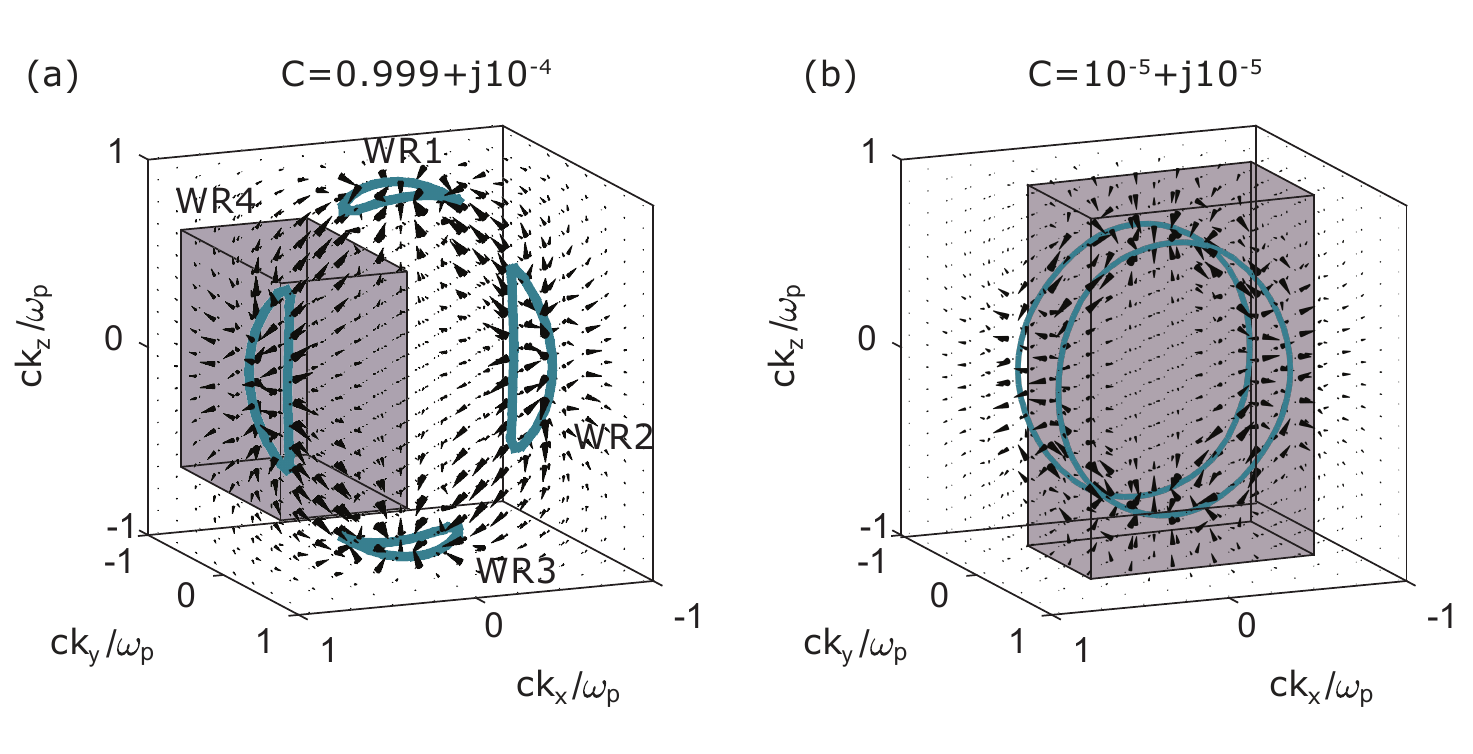}
	\caption{Topological transition and numerical calculation of the topological charge in a chiral Weyl plasmonic material. (a) Weyl exceptional rings and corresponding real Berry curvature for $\alpha_{zz}=0.5$ and $\gamma=0.5\omega_p$. As an example, the complex Berry curvature is numerically integrated over the shaded cuboid surrounding WR4. The numerical calculation results in an approximate topological charge $C=+1$. (b) Weyl exceptional rings and corresponding real Berry curvature for $\alpha_{zz}=0.3$ and $\gamma=0.7\omega_p$. In this regime the four Weyl exceptional rings merge and transform into two exceptional rings carrying no charge. This is confirmed by integrating the complex Berry curvature passing through the shaded cuboid surrounding one of the new exceptional rings.}\label{fig:6}
\end{figure}

We carried out a similar analysis for the chiral plasmonic Weyl material, for which the Weyl degeneracies are not on parallel planes, as discussed above. For a fixed value of chiral coupling $\alpha_{zz}$, we have verified that increasing dissipation results in Weyl exceptional rings of opposite charge intersecting at a critical value of isotropic loss. This behavior is shown in Fig. \ref{fig:5}(b): as $\gamma$ is increased from $0.2\omega_p$ to $0.7\omega_p$ for $\alpha_{zz}=0.5$, the four Weyl exceptional rings, WR1-4, become roughly elliptical and grow in size until they intersect, at the same frequency, for $\gamma=0.7\omega_p$, which results in a topological transition. 
In Fig. \ref{fig:6}, we also show how the exceptional contours, and the corresponding Berry curvature, evolve for losses larger than the critical value. Our numerical calculations show that the new exceptional contours are indeed topologically trivial as they carry zero topological charge.

Decreasing $\alpha_{zz}$ for a fixed $\gamma$ increases the size of the Weyl exceptional rings, which results in a topological transition at a lower critical value of loss. By sweeping $\alpha_{zz}$ and $\gamma$, we plot the phase diagram for this system in Fig. \ref{fig:5}(a). In stark contrast to the magnetized plasma case, the critical value of loss is directly proportional to $\alpha_{zz}$ even when the parameter responsible for the topological nature of the system, in this case $\alpha_{zz}$, is small. Specifically, this fact means that, for small chiral coupling, small dissipative losses in the plasmonic material can result in a topological transition. These findings confirm that, compared to the magnetized plasma realization, the Weyl topological state realized in a chiral plasmonic material is fundamentally more fragile to dissipation. The effect of losses, therefore, should be carefully assessed in any Weyl material that is based on breaking inversion symmetry, instead of time-reversal symmetry. 

\section{Discussion}  

Dissipation is a basic phenomenon of wave physics, with important implications for the operation of classical and quantum devices. Within this context, in this work, we have shown that the impact of dissipation on the topological properties of a three-dimensional Weyl material can be deduced from relevant symmetry arguments, which provides insight into the possibility of dissipation-induced topological transitions. Furthermore, we have investigated nonreciprocal and chiral plasmonic Weyl materials as a relevant model system for continuous topological Weyl media. The physics of plasmonic systems with broken time-reversal or inversion symmetry is incredibly rich, with a plethora of different topological and non-topological degeneracies, including Weyl points, exceptional points and rings, nodal rings, bulk Fermi arcs, etc. Based on this model system, we have shown that Weyl points in a biased plasma transform into Weyl rings of exceptional points with an integer topological charge. In theory, very high values of loss may cause oppositely charged Weyl exceptional rings to come into contact and dissipate their topological charge. However, we have also demonstrated that, even for low cyclotron frequencies, this topological transition requires an unrealistically high value of dissipation. Our findings allows us to conclude that, in the specific case of a nonreciprocal plasma, a topological transition solely due to the presence of losses is possible in theory, but is practically unobservable in a physical system. Conversely, we have shown that chiral plasmonic materials with broken inversion symmetry are much more fragile to the impact of dissipation. Loss-induced topological transitions are possible for weakly chiral materials even for low levels of losses. 

In summary, our results clarify under what conditions dissipation may provide a mechanism for inducing topological transitions in different realizations of electromagnetic Weyl materials. We believe that knowing, a priori, when dissipation may lead to a change in the topological nature of a system, and how the transition may be avoided, are invaluable pieces of information, especially for the applications of three-dimensional topological materials in practical scenarios.

\begin{acknowledgments}
We acknowledge support from the National Science Foundation (NSF) with Grant No. 1741694, and the Air Force Office of Scientific Research with Grant No. FA9550-19-1-0043.
\end{acknowledgments}





%

\end{document}